\documentclass[
	a4paper, 
	10pt, 
	unnumberedsections, 
	twoside, 
]{LTJournalArticle}

\runninghead{} 

\footertext{\textit{RISC-V Summit Europe, Paris, France, 12-15th May 2025}} 

\newcommand{\hwmark}{\textsuperscript{*}}

\begin{document}

\title{Accelerating GenAI Workloads by Enabling RISC-V Microkernel Support in IREE}

\author{%
	Adeel Ahmad, Ahmad Tameem Kamal, Nouman Amir, Bilal Zafar, Saad Bin Nasir
}

\date{\footnotesize\textsuperscript{\textbf{}}10xEngineers}

\renewcommand{\maketitlehookd}{%
\begin{abstract}
This project enables RISC-V microkernel support in IREE, an MLIR-based machine learning compiler and runtime. The approach begins by enabling the lowering of MLIR linalg dialect contraction ops to linalg.mmt4d op for the RISC-V64 target within the IREE pass pipeline, followed by the development of optimized microkernels for RISC-V. The performance gains are compared with upstream IREE and Llama.cpp for the Llama-3.2-1B-Instruct model.
\end{abstract}
}


\maketitle 


\section{Introduction}

IREE (Intermediate Representation Execution Environment)\cite{b1}, an MLIR-based\cite{b2} compiler and runtime, was developed to compile machine learning models for various platforms, such as CPUs, GPUs, and accelerators. IREE accepts ML workload in the form of MLIR code as input, applies classical optimizations like operator fusion and tiling on it, and generates optimized binary for execution. While the compiler-generated code performs reasonably well in most cases, custom kernels perform better in mixed precision calculations. IREE includes a library of optimized microkernels for various CPU and GPU architectures. IREE includes microkernels for x86 and ARM CPUs, however, despite the increasing presence of RISC-V in the AI hardware space, RISC-V microkernels are still missing. This limitation results in poor performance of GenAI models on RISC-V-based hardware. To address this issue, this project enables RISC-V microkernel support in IREE.

\section{Theoretical Framework}

Matrix multiplication is a critical computation in GenAI workloads. IREE uses MLIR linalg dialect contraction ops for matrix multiplication that undergoes tiling and fusion as it progresses through the compilation pipeline. However, tiled matmul has suboptimal performance if the data is not pre-arranged, leading to a high cache miss rate\cite{b3}. To address this, IREE utilizes \textit{\textbf{tensor.pack}} MLIR operation to rearrange the data, ensuring that tiles are stored contiguously in memory before applying \textbf{\textit{linalg.mmt4d}} for optimized computation. The 4-D matrix produced by \textit{linalg.mmt4d} is then converted back to the original layout, using \textbf{\textit{tensor.unpack}} operation.
\begin{enumerate}
    \item \textbf{\textit{tensor.pack}:} It takes a 2-D matrix and converts it into a tiled 4-D matrix in which all tiles are stored contiguously in memory.
    \item \textbf{\textit{linalg.mmt4d}:} It performs matrix multiplication between 4-D left-hand and right-side matrices, produced by the pack operations. The `t` in this stands for the transpose of the right-hand-side matrix.
    \item \textbf{\textit{tensor.unpack}:} It converts the 4-D result matrix produced by mmt4d back to 2-D layout.
\end{enumerate}
These MLIR operations are lowered into calls to microkernels by IREE compilation passes. Even though IREE's architecture\cite{b4} is designed to make it easier for users to integrate arbitrary microkernels, currently it only contains microkernels for pack, unpack, and mmt4d operations, for various precisions, for both x86 and ARM64. For the RISC-V target, we have implemented mmt4d microkernels.

\section{Methodology}
The proposed methodology can be viewed as a two step process:
\begin{enumerate}
    \item  The first part involves enabling transformations of linalg contraction ops to \textit{tensor.pack}, \textit{tensor.unpack} and \textit{linalg.mmt4d} operations. Currently, the transformation of linalg contraction ops to \textit{linalg.mmt4d} (and \textit{tensor.pack}, \textit{tensor.unpack}) is performed within the \textbf{\textit{iree-codegen-materialize-device-encoding}} pass for x86-64 and ARM64 targets. This pass determines the tile sizes for the \textit{M}, \textit{N}, and \textit{K} dimensions of the input matrices based on the target architecture. We modified this pass to enable the transformation of contraction ops into \textit{linalg.mmt4d} and to perform the \textbf{VLEN-aware tiling} for RISC-V64 target. Once this transformation is complete, the \textit{linalg.mmt4d} operation get transformed into calls to the microkernel functions via the subsequent passes. Tile sizes were selected based on the following strategy\cite{b11}:
\begin{enumerate}
    \item \textbf{Prefill}: Tile Size= M,N,K=6,VLEN/8,1
    \item \textbf{Decode}: Tile Size=M,N,K=1,VLEN/4,1
\end{enumerate}
    It was observed that choosing a smaller tile size than these leads to underutilization of hardware registers, while using bigger tile sizes increases register pressure that causes register spills and reloads and degrades performance.
    \newline
    \item The second step includes implementing the microkernel functions. The mmt4d microkernels were implemented for the \textit{f16}x\textit{f16->f32} case, where rhs and lhs operands are of f16 type and the result operand is of type f32. Separate mmt4d microkernels were implemented for LLM's prefill and decode phases, because prefill has GEMM while the decode phase has GEMV computations.
\end{enumerate}
\section{Testing and Performance Benchmarking}
To verify the accuracy of the newly implemented microkernels, we evaluated the Llama-3.2-1B-Instruct model compiled with our microkernels, using our framework built on top of \textbf{LM-Evaluation-Harness}\cite{b5}. The results are summarized in Table 1. The model compiled with 10x-IREE has exactly the same scores as the one obtained from Huggingface.

\begin{table}[h]
\centering
\begin{tabular}{|c|c|c|}
\hline
Benchmark & Huggingface & 10x-IREE \\
\hline
ARC\_c & 59.4\% & 59.4\% \\
\hline
GPQA & 27.2\% & 27.2\% \\
\hline
\end{tabular}
\caption{Evaluation results of the LLaMA-3.2-1B-Instruct model on selected benchmarks. The table compares the performance of two versions of the model: one downloaded from Hugging Face and the other compiled using 10x-IREE.}
\end{table}

For performance benchmarking Llama-3.2-1B-Instruct model was compiled using 10x-IREE, and tokens per second were recorded for prefill and decode phases. The results are summarized in Table 2. For the single-threaded run, we observed 50x gain in decode performance as compared to the upstream IREE. For multi-threaded run, a performance gain of 2x was observed in the prefill phase and 17x was observed in the decode phase.

\begin{table}[h]
\small
\centering
\begin{tabular}{|c|c|c|c|c|}
\hline
Phase                & Threads & Llama.cpp & IREE & 10x-IREE \\
\hline
\multirow{2}{*}{Prefill} & 1 & 0.04 & 0.14 & 0.18 \\
                        & 8 & 0.11 & 0.91 & 1.89 \\
\hline
\multirow{2}{*}{Decode} & 1 & 0.03 & 0.02 & 0.99 \\
                        & 8 & 0.07 & 0.12 & 2.12 \\
\hline
\end{tabular}
\caption{Performance(reported as tokens per second) of the LLaMA-3.2-1B-Instruct model in prefill and decode stages, compiled using llama.cpp, IREE, and 10x-IREE.\hwmark}
\end{table}

\begingroup
\renewcommand{\thefootnote}{*}
\footnotetext{Benchmarking was conducted on a MILK-V Jupiter board featuring 1.66GHz × 8 RISC-V vector cores, with VLEN=256-bits and RVA22 profile.}
\endgroup

\begin{figure}[H]
    \centering
    \includegraphics[width=0.4\textwidth]{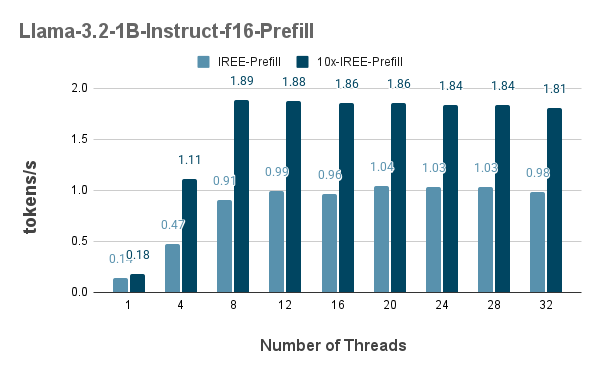}
    \caption{Prefill phase performance comparison between Llama-3.2-1B-Instruct compiled with IREE and 10x-IREE.\hwmark}
    \label{fig:example}
\end{figure}

\begin{figure}[H]
    \centering
    \includegraphics[width=0.4\textwidth]{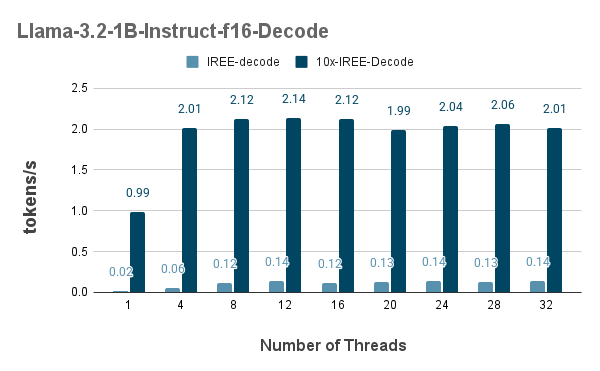}
    \caption{Decode phase performance comparison between Llama-3.2-1B-Instruct compiled with IREE and 10x-IREE.\hwmark}
    \label{fig:example2}
\end{figure}


\bibliographystyle{IEEEtran}
\bibliography{bibliography}
\end{document}